\documentclass{article}
\usepackage{amsmath,amsthm,amssymb,authblk}

\newtheorem{theorem}{Theorem}[section]
\newtheorem{corollary}[theorem]{Corollary}

\newtheorem{lemma}[theorem]{Lemma}
\newtheorem{definition}[theorem]{Definition}

\newtheorem{remark}[theorem]{Remark}
\numberwithin{equation}{section}

\begin{document}
\title{Thresholds of Random Quasi-Abelian Codes}
\author{Yun Fan,\quad Liren Lin\\
\small School of Mathematics and Statistics\\
\small Central China Normal University, Wuhan 430079, China}
\date{}
\maketitle

\insert\footins{\footnotesize{\it Email address}:
yfan@mail.ccnu.edu.cn (Yun Fan); L\_R\_Lin@163.com (Liren Lin).}

\begin{abstract}
For a random quasi-abelian code of rate $r$, it is shown that
the GV-bound is a threshold point:
if $r$ is less than the GV-bound at $\delta$, then the probability
of the relative distance of the random code
being greater than~$\delta$ is almost~$1$;
whereas, if $r$ is bigger than the GV-bound at~$\delta$,
then the probability is almost $0$. As a consequence,
there exist many asymptotically good quasi-abelian codes with any
parameters attaining the GV-bound.\\

\end{abstract}

\section{Introduction}

Random codes play an important role in Informatics, Statistical Physics
and Coding Theory; for example, see \cite{BF}, \cite{MM}.
For a random linear code of rate~$r$ over a finite field $F$ with $q$ elements,
Varshamov \cite{V} and Pierce \cite{P67} showed in fact that
the GV-bound (see~(\ref{GV}) below) is a threshold point:
if $r$ is less than the GV-bound at~$\delta$ where $0<\delta<1-q^{-1}$,
then the probability of the relative distance of the random linear code
being greater than $\delta$ is almost~$1$;
whereas, if~$r$ is bigger than the GV-bound at $\delta$,
then the probability is almost~$0$. Recently, in \cite{FLLSX}
the {\em cumulative distance enumerators} of random codes are introduced
and their thresholds are investigated; as a consequence,
the above threshold of random linear codes is redescribed explicitly
with the parameters $r$ and $\delta$.

By means of random codes, \cite{CPW} showed that,
if 2 is primitive for infinitely many primes
(this is a so-called {\em Artin's conjecture}),
then the asymptotically good binary quasi-cyclic codes exist.
Later, \cite{C} and \cite{K} made big improvements
from different points of view and proved that,
without the Artin's conjecture,
the asymptotically good binary quasi-cyclic codes exist.

For a finite group $G$ of order $m$, any element $\sum_{z\in G}a_zz$
(with $a_z\in F$) of the group algebra $FG$ over the finite field $F$
can be viewed as a word $(a_z)_{z\in G}$ of length $m$ over $F$.
By extension, any element of the free module $(FG)^n$ of rank~$n$
can be viewed as a word of length $mn$.
Any $FG$-submodule $C$ of $(FG)^n$ is called a
{\em quasi-group code} of index $n$.
The code $C$ is just the so-called {\em group code} if $n=1$;
whereas it is just the usual quasi-cyclic code of index $n$ if $G$ is cyclic.
And, $C$ is called a {\em quasi-abelian code} if $G$ is abelian;
see \cite{DR}, \cite{W}.

In 2006, Bazzi and Mitter \cite{BM} constructed a class of
random binary quasi-abelian codes and a class of
random binary dihedral group codes, and showed that the probability
of the parameters of the random codes of any one of the two classes
attaining GV-bound is large;
as a consequence, within the two classes the asymptotically good codes exist.
Soon after, with the similar random method
Mart\'inez-P\'erez and Willems \cite{MW} proved that
self-dual doubly-even binary dihedral group codes are asymptotically good.

We are interested in general random quasi-abelian codes and their thresholds.
Modifying the random linear code ensemble in Shannon's Information Theory
(cf. \cite[ch.6]{MM}), in Section 2 we construct the general
{\em random quasi-abelian code ensemble}, and state our main theorem,
see Theorem \ref{main theorem} below, which asserts that
the GV-bound is still a threshold point, i.e.
the probability of the relative distance of the random code of the ensemble
being greater than a given $\delta$ is almost $1$
if the parameters are below the GV-bound; whereas, the probability is almost $0$
if the parameters are beyond the GV-bound.
The Varshamov-Pierce's threshold for random linear codes mentioned above
is the special case of our main theorem by taking the finite group to be trivial.

The proof of the main theorem consists of three parts.
In Section 3, we extend a result on weights of so-called {\em balanced codes};
this result appeared in \cite{M74}, \cite{P85} and \cite{S86} in
a binary version,  which played a key role in  \cite{BM} and \cite{MW}.
We generalize it to any $q$-ary version,
see Theorem \ref{a-weight balanced} below, so that we can treat any $q$-ary codes.
Theorem \ref{a-weight balanced} has independent significance;
for example,  from it
quite a part of~\cite{BM} can be extended to any $q$-ary case.

In Section 4, a threshold of the expectation of the cumulative weight enumerator
of the random code of the ensemble is obtained in Theorem \ref{T-CWE} below,
from which the first part (``below the GV-bound'') of the main theorem
follows immediately.

In Section 5, we prove the second part (``beyond the GV-bound'')
of the main theorem by estimating the second moment of
the cumulative weight enumerator of the random code of the ensemble.

From the random quasi-abelian code ensemble and the main theorem,
in Section 6, we draw the random quasi-abelian codes of given rate $r$
and describe their thresholds; in particular,
for any finite abelian group, for any $r$ and $\delta$ attaining the GV-bound,
there is a series of quasi-abelian codes such that the limit of their rates and
the limit of their relative distances are equal to $r$ and $\delta$ respectively.

In this paper, $h_q(x)=x\log_q(q-1)-x\log_q x-(1-x)\log_q(1-x)$
with the convention that $0\log_q0=0$,
the function $h_q(x)$ is called the {\em $q$-ary entropy}
(different from the entropy with base $q$ in Informatics, see \cite[\S2.1]{CT}); 
and let
\begin{equation}\label{GV}
 g_q(x)=1-h_q(x)=1-x\log_q(q-1)
   +x\log_q x+(1-x)\log_q(1-x),
\end{equation}
which is the {\em $q$-ary asymptotic Gilbert-Varshamov bound},
or {\em GV-bound} in short;
note that $g_q(x)$ for $x\in[0,1]$
is a convex function and has a unique zero point at $x=1-q^{-1}$,
hence $g_q(x)$ is a strictly decreasing function
for $x\in[0,1-q^{-1}]$; see \cite[\S2.10.6]{HP}.
About fundamentals on coding theory and group theory,
please refer to \cite{HP} and \cite{HB} respectively.

\section{Random quasi-abelian code ensembles}

In this paper {\em we always assume} that
$F$ is a finite field with cardinality $|F|=q=p^e$ where $p$ is a prime,
and $G$ is a finite abelian group of order $|G|=m$.

By $FG=\big\{\sum_{z\in G}a_z z~\big|~ a_z\in F\big\}$
we denote the group algebra of $G$ over $F$.
Each element $a=\sum_{z\in G}a_z z$ of $FG$ is viewed as
a word $(a_z)_{z\in G}$ of length $m$ over~$F$,
and ${\rm w}(a)={\rm w}((a_z)_{z\in G})$ stands for the usual
Hamming weight of the word $(a_z)_{z\in G}$. In this way,
$a=\sum_{z\in G}a_z z\in FG$ and word $(a_z)_{z\in G}\in F^m$
are identified with each other; but note that
for $a,b\in FG$ we have the product $ab$ in the algebra~$FG$.

Let $n$ be any positive integer. We consider the free $FG$-module of rank $n$:
$$(FG)^n=\big\{{\bf a}=(a_1,\cdots,a_n)
 ~\big|~a_i\in FG,~ i=1,\cdots,n\big\}.$$
Each element ${\bf a}=(a_1,\cdots,a_n)\in (FG)^n$ is identified
with a concatenated word
$\big((a_{1z})_{z\in G},\cdots,(a_{nz})_{z\in G}\big)$
of length $mn$ over $F$, thus the Hamming weight
${\rm w}({\bf a})={\rm w}(a_1,\cdots,a_n)={\rm w}(a_1)+\cdots+{\rm w}(a_n)$.
As mentioned in Introduction,
any submodule $C$ of the $FG$-module $(FG)^n$ is said to be
a {\em quasi-abelian code} of $G$ over~$F$
(or {\em quasi-$FG$ code} more precisely) with index~$n$.
In particular, it is just the usual {\em abelian code} if $n=1$;
whereas, it is just the usual {\em quasi-cyclic code} with index $n$
if $G$ is cyclic.

We always take the following parameters:
\begin{equation}\label{parameters}
r\in(0,1),\qquad \delta\in(0,\delta_0)
~~ {\rm where}~~ \delta_0=1-q^{-1},
\end{equation}
and set $k=[rn]$, 
the integer nearest to $rn$.
We consider the set of $k\times n$ matrices over $FG$:
\begin{equation}\label{ensemble}
(FG)^{k\times n}
=\left\{A=\begin{pmatrix}a_{11}&\cdots&a_{1n}\\
 \cdots&\cdots&\cdots\\ a_{k1}&\cdots&a_{kn}\end{pmatrix}
 ~\Bigg|~ a_{ij}\in FG\right\},
\end{equation}
which is viewed as a probability space
with equiprobability. Following a notation in Shannon's
information theory, we call this
probability space the {\em random quasi-abelian code ensemble}.
In particular, if $G=1$  is trivial then $(FG)^{k\times n}=F^{k\times n}$
is just the usual  {\em random linear code ensemble};
cf. \cite[ch.6]{MM}.

Take $A=(a_{ij})_{k\times n}\in (FG)^{k\times n}$, i.e.
$A$ is a random $k\times n$ matrix over $FG$.
We write $A=(A_1,\cdots,A_n)$ with $A_j=(a_{1j},\cdots,a_{kj})^T$
being the $j$'th column of the matrix $A$,
where the superscript ``$T$'' stands for the transpose.
Then we have a {\em random quasi-abelian code} $C_A$ of index $n$ as follows:
\begin{equation}\label{C_A}
 C_A=\Big\{{\bf b}A=\big({\bf b}A_1,~\cdots,~{\bf b}A_n\big)
 ~\Big|~{\bf b}=(b_1,\cdots,b_k)\in(FG)^k\Big\},
\end{equation}
where ${\bf b}A_j=b_1a_{ij}+\cdots+b_ka_{kj}\in FG$.
Note that the rate $R(C_A)=\frac{\dim C_A}{mn}$.
It is obvious that $R(C_A)\le \frac{k}{n}\approx r$,
and $R(C_A)=\frac{k}{n}$ if and only if the $FG$-rank of~$A$ is equal to $k$;
so, we can get the random quasi-abelian codes of rate $r$ from the ensemble,
see Section 6 below.
About the rank of a matrix over a ring, please see \cite[\S2]{FLL},
or related refs such as \cite{Mc}.

By $\Delta(C_A)$ we denote the relative distance of
the random quasi-abelian code~$C_A$, i.e.
$\Delta(C_A)=\frac{{\rm w}(C_A)}{mn}$, where ${\rm w}(C_A)$ denotes
the minimum weight of~$C_A$. Then $\Delta(C_A)$ is
a random variable over the probability space~$(FG)^{k\times n}$.
We consider the asymptotic property (with $n\to\infty$) of
$\Pr\big(\Delta(C_A)>\delta\big)$ which stands for the probability that
$\Delta(C_A)>\delta$,
and state our main theorem.

\begin{theorem}\label{main theorem}
Let notations be as in (\ref{parameters}), (\ref{ensemble}) and (\ref{C_A}). Then
$$\lim\limits_{n\to\infty}\Pr\big(\Delta(C_A)>\delta\big)=
\begin{cases}
    1, & \mbox{if~ $r<g_q(\delta)$;} \\
    0, & \mbox{if~ $r>g_q(\delta)$;}
\end{cases}
$$
and both the limits converge exponentially.
\end{theorem}

If $G=1$ is trivial, then $FG=F$ is just the finite field $F$ and
the theorem exhibits just the threshold of random linear codes
obtained by Vasharmov \cite{V} and Pierce \cite{P67}
(cf, \cite[Corollary 3.2]{FLLSX}), as mentioned in Introduction.

The key idea for the proof of the theorem is to estimate the first moment
(i.e. the expectation) and the second moment of the cumulative weight
enumerator of the random code $C_A$, so that we can bound
$\Pr\big(\Delta(C_A)>\delta\big)$ suitably;
for estimating the moments we need a result on weights of balanced
codes which appeared in references, as we've seen so far, only in binary version,
so we extend it to $q$-ary version first.
Thus, as we mentioned in Introduction,
the proof of the main theorem will be completed in Sections 3, 4 and 5.

\section{The weights of balanced codes}

Let $I=\{1,2,\cdots,n\}$ be an index set;
let $F^I=F^n$ be the set of all words over~$F$ of length $n$.
For any subset $I'=\{i_1,\cdots,i_d\}$ of $I$
with $1\le i_1<\cdots<i_d\le n$, we have a projection
$\rho'$ from $F^I$ to $F^{I'}$ as follows:
$\rho'({\bf a})=(a_{i_1},\cdots,a_{i_d})\in F^{I'}$
for any ${\bf a}=(a_{1},\cdots,a_{n})\in F^{I}$.

\begin{definition}\label{def balanced}\rm
 Let $C\subseteq F^n=F^I$.
If there are subsets $I_1,\cdots,I_s$ (with repetition allowed)
of the index set $I$ with every cardinality $|I_j|=d$
and an integer $t$ such that
\begin{itemize}
\item[(i)] for any index $i\in I$, the number of the subscripts
$j$ satisfying that $i\in I_j$ is equal to $t$;

\item[(ii)] for any $j=1,\cdots,s$,
the projection $\rho_j:F^I\to F^{I_j}$ maps $C$ bijectively
onto $F^{I_j}$;
\end{itemize}
then we say that $C$ is a {\em balanced code} of $F^n$ with
{\em information length} $d$, and $I_1,\cdots,I_s$ form
a {\em balanced system of information index sets} of $C$.
\end{definition}

\begin{remark}\label{g-code balanced}\rm
For example, any group code $C$ (i.e. any ideal) of
the group algebra $FG$ is a balanced code, see \cite{BM};
similarly, any coset $a+C$ for $a\in FG$ is a balanced code too.
\end{remark}

For any word ${\bf a}=(a_1,\cdots,a_n)\in F^n$,
the fraction ${\rm w}({\bf a})/n$
is called the {\em relative weight} of ${\bf a}$.
The following is a generalization of a result in
\cite{M74}, \cite{P85} and~\cite{S86}, where only the binary case is considered.

\begin{theorem}\label{a-weight balanced}
Let $C$ be a balanced code of $F^n$ with information length $d$
and $B$ be a non-empty subset of $C$, and let
$\omega=\frac{\sum_{{\bf b}\in B}{\rm w}({\bf b})}{n|B|}$
(the average relative weight of $B$). If $0\le\omega\le 1-q^{-1}$, then
\begin{equation}\label{weight balanced}
 |B|\le q^{d h_q(\omega)}.\end{equation}
\end{theorem}

Before proving the theorem, we show two corollaries.

\begin{corollary}\label{m-weight balanced}
Let $C$ be a balanced code of $F^n$ with information length $d$,
let $C^{\le\delta}$ be the set of the codewords of $C$
which relative weight are at most $\delta$.
If $0\le\delta\le 1-q^{-1}$, then
$
 |C^{\le\delta}|\le q^{dh_q(\delta)}.
$
\end{corollary}

{\bf Proof.}~ The average relative weight
of $C^{\le\delta}$ is at most $\delta$,
and $h_q(x)$ is an increasing function in $[0,1-q^{-1}]$. \qed

\medskip
For $C\subseteq F^n$,  the Cartesian product of $n'$ copies of $C$
in $(F^n)^{n'}$ is as follows:
\begin{equation}\label{product}
 C^{n'}=\big\{({\bf c}_1,\cdots,{\bf c}_{n'})
  \;\big|\; {\bf c}_i\in C,~i=1,\cdots,n'\big\}.
\end{equation}

\begin{corollary}\label{product balanced}
Let $C$ be a balanced code of $F^n$
with information length $d$. Then the product code $C^{n'}$
is a balanced code of $F^{nn'}$ with information length $dn'$;
in particular, if $0\le\delta\le 1-q^{-1}$ then
$
\big|(C^{n'})^{\le\delta}\big|\le q^{dn'h_q(\delta)}.
$
\end{corollary}

\smallskip{\bf Proof.}~ Assume that
the subsets $I_1,\cdots,I_s$ of the index set $I=\{1,\cdots,n\}$
form a balanced system of information index sets of~$C$.
We write the index set of the product code $C^{n'}$ as:
$$ I^{n'}=\big\{1^{(1)},\cdots,n^{(1)},~\cdots,~
   1^{(n')},\cdots,n^{(n')}\big\}.$$
For each $I_j=\{j_1,\cdots,j_d\}$,
we can form a subset $I_j^{n'}$ of $I^{n'}$ by
concatenating $n'$ copies of $I_j$ as follows:
$$ I_j^{n'}=\big\{j_1^{(1)},\cdots,j_d^{(1)},~\cdots,~
  j_1^{(n')},\cdots,j_d^{(n')}\big\}.  $$
Then it is easy to check that
$I_1^{n'},~\cdots,~I_s^{n'}$
form a balanced system of information index sets
of the product code $C^{n'}$. \qed

\medskip The rest of this section contributes to the proof of the theorem.

\medskip
{\bf Proof of Theorem \ref{a-weight balanced}.}~
First we assume that $d=n$, i.e. $C=F^n$
which is of course balanced (with $s=1$, $I_1=I$ and $t=1$),
and prove the inequality (\ref{weight balanced});
this is a key step of the proof.

Set $M=|B|$.
Consider $B$ as a probability space with equiprobability.
Each ${\bf b}\in B$ is an $n$-tuple: ${\bf b}=(b_1,\cdots,b_n)$.
For each index $i$, $1\le i\le n$, we have a random
variable $X_i$ defined over the probability space $B$
and taking values in $F$ as follows: $X_i({\bf b})=b_i$;
hence we have a discrete distribution function
$p_i(a)=\Pr(X_i=a)$ for $a\in F$; we write the distribution as:
$$ p_i=\big(p_i(a)\big)_{a\in F}\,,\qquad
 i=1,\cdots,n. $$
Set
$$
 p=\frac{p_1+\cdots+p_n}{n};
$$
then $p$ is a distribution function.
Denote $F^*=F\backslash\{0\}$ (which denotes the difference set).
It is obvious that
\begin{equation}\label{omega}
 \omega=\sum_{a\in F^*}p(a)
 =\sum_{a\in F^*}\sum_{i=1}^n \frac{p_i(a)}{n};
\end{equation}
hence we also have that
\begin{equation}\label{1-omega}
1-\omega=p(0)=\sum_{i=1}^n \frac{p_i(0)}{n}.
\end{equation}
Consider the random $n$-tuple ${\bf X}=(X_1,\cdots,X_n)$ and its
entropy with base $q$:
\begin{eqnarray*}H_q({\bf X})=H_q(X_1,\cdots,X_n)
=\sum_{{\bf a}\in F^n}-\Pr({\bf X}={\bf a})\log_q\Pr({\bf X}={\bf a}).
\end{eqnarray*}
For any ${\bf a}=(a_1,\cdots,a_n)\in F^n$, by the definition of
the random variables $X_i$'s, we have
$$
 \Pr({\bf X}={\bf a})
 =\begin{cases}\frac{1}{M}, & {\bf a}\in B;\\
 0, &{\bf a}\notin B.\end{cases}
$$
So we get
\begin{equation}\label{log M}
 H_q({\bf X})=H_q(X_1,\cdots,X_n)=\log_q M.
\end{equation}
On the other hand, by an inequality for entropy of joint distribution
(see \cite[Theorem 2.6.6]{CT}), we have
\begin{equation*}
H_q(X_1,\cdots,X_n)
\le H_q(X_1)+\cdots+H_q(X_n)
    = \sum_{i=1}^n\sum_{a\in F}-p_i(a)\log_q p_i(a);
\end{equation*}
so
\begin{eqnarray*}
H_q({\bf X})\le
\left(\sum_{i=1}^n -p_i(0)\log_q p_i(0)\right)+\left(
\sum_{i=1}^n \sum_{a\in F^*}-p_i(a)\log_q p_i(a)\right).
\end{eqnarray*}
Since $-x\log_q x$ is a concave function, for the second bracket of the right 
hand side of the above inequality we get (with the help of Eqn (\ref{omega}))
\begin{eqnarray*}
\frac{\sum\limits_{i=1}^n\sum\limits_{a\in F^*}
  -p_i(a)\log_q p_i(a)}{n(q-1)}
&\le&
-\frac{\sum\limits_{i=1}^n\sum\limits_{a\in F^*}p_i(a)}{n(q-1)}
 \log_q\frac{\sum\limits_{i=1}^n\sum\limits_{a\in F^*}
  p_i(a)}{n(q-1)}\\
&=&-\frac{\omega}{q-1}\log_q\frac{\omega}{q-1};
\end{eqnarray*}
that is
$$
\sum_{i=1}^n\sum_{a\in F^*}-p_i(a)\log_q p_i(a)\le
n\Big(\omega\log_q(q-1)-\omega\log_q\omega\Big).
$$
Similarly, with the help of Eqn (\ref{1-omega}) we can obtain
$$
\sum_{i=1}^n -p_i(0)\log_q p_i(0)\le -n(1-\omega)\log_q(1-\omega).
$$
Thus we get
$$
H_q({\bf X})\le n\Big(\omega\log_q(q-1)
-\omega\log_q\omega-(1-\omega)\log_q(1-\omega)\Big)
=nh_q(\omega).
$$
Combining it with Eqn (\ref{log M}), we obtain that
\begin{equation*}
\log_q|B|=\log_q M\le n h_q(\omega).
\end{equation*}
which is just the inequality (\ref{weight balanced}) since
we have assumed that $d=n$.

Next we turn to the general case. That is, there are subsets
$I_1,\cdots,I_s$ of the index set $I=\{1,2,\cdots,n\}$ with each
$|I_j|=d$ such that any index $i\in I$ appears in exactly $t$
members of the $s$ subsets $I_1,\cdots,I_s$; in particular, we have
\begin{equation}\label{tn=sd} tn=sd.\end{equation}
Set $|B|=M$ again. For each $I_j$, by $\rho_j$ we denote the projection
from $F^n=F^I$ onto $F^{I_j}$; then $|\rho_j(B)|=M$.

Let $\hat I$ be the disjoint union of $I_1,\cdots,I_s$
(though they may be not disjoint),
so $|\hat I|=sd$, and $F^{\hat I}=F^{I_1}\times\cdots\times F^{I_s}$
is the product of $F^{I_j}$ for $j=1,\cdots,s$, i.e.
the words of $F^{\hat I}$ are the concatenations
of the words of $F^{I_j}$ for $j=1,\cdots,s$:
$$F^{\hat I}=\Big\{({\bf a}_1,\cdots,{\bf a}_s)~\Big|~
 {\bf a}_j\in F^{I_j},~ j=1,\cdots,s \Big\}.$$
Consider the following subset of $F^{\hat I}$:
$$
\hat B=\rho_1(B)\times\cdots\times\rho_s(B)=
\Big\{\big(\rho_1({\bf b}_{1}),\cdots,\rho_s({\bf b}_{s})\big)~\Big|~
 {\bf b}_1,\cdots,{\bf b}_s\in B\Big\}.
$$
Since $|\rho_j(B)|=M$ for $j=1,\cdots,s$, we see that
\begin{equation}\label{|B|} |\hat B|=M^s. \end{equation}
Set
$\widehat{{\rm w}(\hat B)}=\sum_{\hat{\bf b}\in\hat B}{\rm w}(\hat{\bf b})$,
which can be computed as follows:
\begin{eqnarray*}
\widehat{{\rm w}(\hat B)}&=&\sum_{{\bf b}_1,\cdots,{\bf b}_s\in B}
 {\rm w}\big(\rho_1\big({\bf b}_1),\cdots,\rho_s({\bf b}_s)\big)\\
&=&
\sum_{{\bf b}_1,\cdots,{\bf b}_s\in B}~\sum_{j=1}^s
{\rm w}\big(\rho_j\big({\bf b}_j)\big)
=\sum_{j=1}^s~\sum_{{\bf b}_1,\cdots,{\bf b}_s\in B}
{\rm w}\big(\rho_j\big({\bf b}_j)\big).
\end{eqnarray*}
For $j=1$ we have that
\begin{eqnarray*}
\sum_{{\bf b}_1,\cdots,{\bf b}_s\in B}
{\rm w}\big(\rho_1\big({\bf b}_1)\big)
&=&\sum_{{\bf b}_1\in B}~\sum_{{\bf b}_2,\cdots,{\bf b}_s\in B}
{\rm w}\big(\rho_1\big({\bf b}_1)\big)\\
&=&\sum_{{\bf b}_1\in B}M^{s-1}{\rm w}
   \big(\rho_1\big({\bf b}_1)\big)
= M^{s-1}\sum_{{\bf b}\in B}{\rm w}\big(\rho_1\big({\bf b})\big).
\end{eqnarray*}
Similarly, $\sum_{{\bf b}_1,\cdots,{\bf b}_s\in B}
{\rm w}\big(\rho_j\big({\bf b}_j)\big)
=M^{s-1}\sum_{{\bf b}\in B}{\rm w}\big(\rho_j\big({\bf b})\big)$. So
\begin{eqnarray*}
\widehat{{\rm w}(\hat B)}=\sum_{j=1}^s
M^{s-1}\sum_{{\bf b}\in B}{\rm w}\big(\rho_j\big({\bf b})\big)
=M^{s-1}\sum_{{\bf b}\in B}\sum_{j=1}^s
{\rm w}\big(\rho_j\big({\bf b})\big).
\end{eqnarray*}
By (i) of Definition \ref{def balanced}, we have
$$\sum_{j=1}^s{\rm w}\big(\rho_j\big({\bf b})\big)=
{\rm w}\big(\rho_1\big({\bf b}),\cdots,\rho_s({\bf b})\big)
=t{\rm w}({\bf b}).
$$
Recalling that $\omega=\frac{\sum_{{\bf b}\in B}{\rm w}({\bf b})}{nM}$,
we obtain that
\begin{eqnarray*}
\widehat{{\rm w}(\hat B)}
=M^{s-1}t\sum_{{\bf b}\in B}{\rm w}({\bf b})
=M^{s-1}t\omega nM=\omega tnM^s.
\end{eqnarray*}
By Eqns (\ref{tn=sd}) and (\ref{|B|}), we compute
the average relative weight of $\hat B$ as follows:
$$
\widehat{{\rm w}(\hat B)}\Big/sdM^s=\omega tnM^s\big/sdM^s=\omega.
$$
Applying the conclusion proved in the first step
(i.e. the case ``$d=n$'') to the subset $\hat B$ of $F^{\hat I}$,
we obtain that $M^s=|\hat B|\le q^{sd h_q(\omega)}$;
in other words,
$$|B|=M\le q^{d h_q(\omega)}.$$
Theorem \ref{a-weight balanced} is proved.
\qed

\section{Cumulative weight enumerators of $C_A$}

We keep the notations in (\ref{parameters}), (\ref{ensemble}) and (\ref{C_A}),
and further set
\begin{equation}\label{c-enumerator}
\hat{\cal N}_{C_A}(\delta)=\Big|\big\{
 {\bf b}\in(FG)^k~\big|~1\le{\rm w}({\bf b}A)\le mn\delta\big\}\Big|,
\end{equation}
which is a non-negative integral random variable defined
over the probability space $(FG)^{k\times n}$.
Obviously, $\hat{\cal N}_{C_A}(\delta)$
stands for the number of such elements ${\bf b}$ of $(FG)^k$ that
${\bf b}A$ is a non-zero codewords of $C_A$ with
relative weights at most~$\delta$; so we call it
the {\em cumulative weight enumerator} of the random code $C_A$;
in particular (cf. \cite[\S3]{FLLSX}),
\begin{equation}\label{X-Delta}
 \hat{\cal N}_{C_A}(\delta)\ge 1 ~\iff~ \Delta(C_A)\le\delta\,.
\end{equation}
We are concerned with the asymptotic behavior of the expectation
${\rm E}\big(\hat{\cal N}_{C_A}(\delta)\big)$.
The following is the main result of this section.

\begin{theorem}\label{T-CWE} Let notation be as in (\ref{parameters}),
(\ref{ensemble}), (\ref{C_A}) and (\ref{c-enumerator}). Then
$$\lim\limits_{n\to\infty}{\rm E}
  \big(\hat{\cal N}_{C_A}(\delta)\big)
  =\begin{cases}0, & r<g_q(\delta);\\ \infty, & r>g_q(\delta);
\end{cases}$$
and both the limits converge exponentially.
\end{theorem}

Before proving the theorem, we show that the first part of
Theorem \ref{main theorem} is an immediate consequence of the first part
of the above theorem.

\begin{corollary}\label{1'st part}  If~ $r<g_q(\delta)$ then
$\lim\limits_{n\to\infty}\Pr\big(\Delta(C_A)>\delta\big)=1$
and the convergence speed is exponential.
\end{corollary}

{\bf Proof.}~ By Eqn (\ref{X-Delta}),
Markov's inequality (see \cite[Theorem 3.1]{MU})
and the first part of Theorem \ref{T-CWE}, we have
$$\lim\limits_{n\to\infty}\Pr\big(\Delta(C_A)\le\delta\big)
  =\lim\limits_{n\to\infty}
    \Pr\big(\hat{\cal N}_{C_A}(\delta)\ge 1\big)
  \le\lim\limits_{n\to\infty}
   {\rm E}\big(\hat{\cal N}_{C_A}(\delta)\big)=0. \qed
$$

To prove Theorem \ref{T-CWE} (and Theorem \ref{main theorem} also),
a key step is to write $\hat{\cal N}_{C_A}(\delta)$ as a sum of
Bernoulli random variables.

For every ${\bf b}\in(FG)^k$ we define a Bernoulli random variable
over the probability space $(FG)^{k\times n}$:
$$
 X_{\bf b}=
 \begin{cases} 1, & {\rm if}~~ 1\le{\rm w}({\bf b}A)\le mn\delta;\\
 0, & {\rm otherwise}. \end{cases}
$$
Set $X=\sum_{{\bf b}\in(FG)^k}X_{\bf b}$. It is obvious that $X_{\bf 0}=0$ and
\begin{equation}\label{X=sum}
\hat{\cal N}_{C_A}(\delta)=\sum_{{\bf b}\in(FG)^k}X_{\bf b}=X.
\end{equation}

Fixing any ${\bf b}=(b_1,\dots,b_k)\in (FG)^k$,
we have an $FG$-homomorphism induced by ${\bf b}$ as follows:
\begin{equation}\label{beta_b}
\begin{array}{crcl}
 \beta_{\bf b}: & (FG)^{k\times n} & \longrightarrow & (FG)^n, \\
  &A & \longmapsto &
  {\bf b}A=\big({\bf b}A_1,\cdots,{\bf b}A_n\big).
\end{array}
\end{equation}
For each $j$,  ${\bf b}A_j=b_1a_{1j}+\cdots+b_ka_{kj}$;
so the set of ${\bf b}A_j$ with $A_j$ running over $(FG)^k$
is an ideal of $FG$ generated by $b_1,\cdots,b_k$,
we denote it by $I_{\bf b}$:
\begin{equation*}
 I_{\bf b}=FGb_1+\cdots+ FGb_k, \quad{\rm for}~~
  {\bf b}=(b_1,\cdots,b_k)\in (FG)^k;
\end{equation*}
and denote $d_{\bf b}=\dim I_{\bf b}$.
Thus, the image of $\beta_{\bf b}$ is the product code
$I_{\bf b}^n\subseteq (FG)^n$, and $\dim I_{\bf b}^n=d_{\bf b}n$.

Since $\beta_{\bf b}$ is an $FG$-homomorphism,
the number of the pre-images in $(FG)^{k\times n}$ of every
${\bf a}\in I_{\bf b}^n$ is equal to $\frac{q^{mkn}}{q^{d_{\bf b}n}}$,
which is independent of the choice of~${\bf a}$.
And, by Remark \ref{g-code balanced} and Corollary \ref{product balanced},
we have $\big|(I_{\bf b}^n)^{\le\delta}\big|\le q^{d_{\bf b}nh_q(\delta)}$. So
\begin{equation*}
{\rm E}(X_{\bf b})=\Pr\big(1\le{\rm w}({\bf b}A)\le mn\delta\big)
\le\frac{q^{d_{\bf b}nh_q(\delta)}-1}{q^{d_{\bf b}n}};
\end{equation*}
that is
\begin{equation}\label{estimate EX_b}
{\rm E}(X_{\bf b})\le q^{-d_{\bf b}ng_q(\delta)}-q^{-d_{\bf b}n},\qquad
\forall~~{\bf b}\in(FG)^k.
\end{equation}

For any ideal $I$ of $FG$, we denote $d_I=\dim I$ and set
\begin{equation}\label{I^k*}
  I^{k*}=\big\{{\bf b}\in I^k\;\big|\; I_{\bf b}=I\big\};
\end{equation}
in particular, $(FG)^{k*}=\{{\bf b}\in(FG)^k\mid I_{\bf b}=FG\}$.
Obviously, we have a disjoint union
$(FG)^k=\bigcup\limits_{I\le FG}I^{k*}$,
where the subscript ``$I\le FG$''  means that
$I$ runs over the ideals of $FG$.
Thus, by the linearity of expectation, we get
\begin{equation}\label{decomp EXI}
{\rm E}(X)={\rm E}\left(\sum_{{\bf b}\in(FG)^k}X_{\bf b}\right)=
\sum_{0\ne I\le FG}~\sum_{{\bf b}\in I^{k*}}{\rm E}(X_{\bf b}).
\end{equation}

To get a lower bound of ${\rm  E}(X_{\bf b})$ for ${\bf b}\in(FG)^{k*}$,
we recall an estimation of a partial sum of binomials:
\begin{equation}\label{estimate PSB}
 q^{nh_q(k/n)-\frac{1}{2}\log_q n}\le
  \sum_{i=1}^k\binom{n}{i}(q-1)^i\le q^{nh_q(k/n)},
\end{equation}
see \cite[Eqn(2.3)]{FLLSX}. One can also check
the upper bound of (\ref{estimate PSB}) from Corollary~\ref{m-weight balanced}
(by taking $C=F^n$, i.e. $k=n$ in the corollary), and check the lower bound
by the argument in \cite[Lemma 9.2]{MU}.

Let ${\bf b_1}\in(FG)^{k*}$, i.e. $I_{\bf b_1}=FG$;
then the image of $\beta_{\bf b_1}$ in (\ref{beta_b}) is just the whole space
$(FG)^n\cong F^{mn}$, so
\begin{equation}\label{EXb*}
{\rm E}(X_{\bf b_1})=
\Pr\Big(1\le{\rm w}({\bf b_1}A)\le mn\delta\Big)
=\frac{\big|(F^{mn})^{\le\delta}\big|-1}{\big|F^{mn}\big|};
\end{equation}
in particular,
\begin{equation}\label{EXb1=EXb2}
 {\rm E}(X_{\bf b_1})={\rm E}(X_{\bf b_2}),\qquad
  \forall~~{\bf b_1}, {\bf b_2}\in(FG)^{k*}.
\end{equation}
Further, since
$\big|(F^{mn})^{\le\delta}\big|=\sum_{i=1}^{mn\delta}\binom{mn}{i}(q-1)^i$
and $\big|F^{mn}\big|=q^{mn}$; by the inequality (\ref{estimate PSB}) we get that
\begin{equation*}
{\rm E}(X_{\bf b_1})\ge q^{-mng_q(\delta)-\frac{1}{2}\log_q(mn)}-q^{-mn},
 \qquad \forall~~ {\bf b_1}\in(FG)^{k*}.
\end{equation*}
Moreover, since $|(FG)^{k*}|=\frac{|(FG)^{k*}|}{|(FG)^k|}\cdot q^{mk}$,
for ${\bf b_1}\in(FG)^{k*}$ we have
\begin{equation}\label{k*E}
|(FG)^{k*}|\cdot{\rm E}(X_{\bf b_1})\ge
\frac{|(FG)^{k*}|}{|(FG)^k|}\Big(
q^{mn\big(\frac{k}{n}-g_q(\delta)\big)-\frac{1}{2}\log_q(mn)}
-q^{-mn(1-\frac{k}{n})}\Big).
\end{equation}

Recalling from (\ref{X=sum}) that $\hat{\cal N}_{C_A}(\delta)=X$,
we show a proof of Theorem \ref{T-CWE}.

\smallskip{\bf Proof of Theorem \ref{T-CWE}.}~

Now we assume that $r<g_q(\delta)$. Since $k=[rn]$,
there is a positive number $\gamma$ such that
for large enough $n$ we have $\frac{k}{n}-g_q(\delta)<-\gamma$.
For any ${\bf b}\in I^{k*}$ as above, since $d_{\bf b}= d_I$,
from Eqn (\ref{estimate EX_b}) we have
${\rm E}(X_{\bf b})\le q^{-nd_Ig_q(\delta)}$.
Further, because $|I^{k*}|\le |I^k|=q^{d_I k}$, we get that
\begin{equation*}
\sum_{{\bf b}\in I^{k*}}{\rm E}(X_{\bf b})
\le q^{d_Ik}q^{-nd_Ig_q(\delta)}
=q^{nd_I\big(\frac{k}{n}-g_q(\delta)\big)}
< q^{-\gamma d_I n},
\end{equation*}
and the right hand side is exponentially convergent to $0$ as $n\to\infty$.
Note that $FG$ has only finitely many ideals,
by Eqn (\ref{decomp EXI}) we obtain that
$$
\lim\limits_{n\to\infty}{\rm E}\big(\hat{\cal N}_{C_A}(\delta)\big)
=\lim\limits_{n\to\infty}{\rm E}(X)
=\sum_{0\ne I\le FG}\lim\limits_{n\to\infty}\sum_{{\bf b}\in I^{k*}}
{\rm E}(X_{\bf b}) =0.
$$

In the following {\em we assume that $r>g_q(\delta)$}. Since $k=[rn]$,
there is a positive number $\gamma$ such that
for large enough $n$ we have $\frac{k}{n}-g_q(\delta)>\gamma$
and $1-\frac{k}{n}>\gamma$.
Fixing a ${\bf b_1}\in(FG)^{k*}$,
from Eqn (\ref{decomp EXI}) and the Eqn (\ref{EXb1=EXb2}) we have:
$$
{\rm E}(X)\ge\sum\limits_{{\bf b}\in(FG)^{k*}}{\rm E}(X_{\bf b})
=|(FG)^{k*}|\cdot{\rm E}(X_{\bf b_1}).
$$
Since $k\to\infty$ as $n\to\infty$, by Lemma \ref{|FGk*|} below,
we have $\lim\limits_{n\to\infty}\frac{|(FG)^{k*}|}{|(FG)^k|}=1$;
so, by the inequality (\ref{k*E}),
we obtain the following exponentially convergent limit:
$$
\lim\limits_{n\to\infty}{\rm E}(X)
>\lim\limits_{n\to\infty}\frac{|(FG)^{k*}|}{|(FG)^k|}
\Big( q^{mn\gamma-\frac{1}{2}\log_q(mn)}-q^{-mn\gamma}\Big)
=\infty.
$$

The proof of Theorem \ref{T-CWE} is finished. \qed

\begin{lemma}\label{|FGk*|} Assume that
$m=|G|=p^{\mu}m'$ with $m'$ coprime to $p$,
$G$ has~$h$ irreducible characters over $F$
with degree $d_1,\cdots,d_h$ respectively,
and $(FG)^{k*}$ is defined as in (\ref{I^k*}). Then the cardinality
\begin{equation}\label{FGk*}
 |(FG)^{k*}|=\prod_{j=1}^h q^{(p^{\mu}-1)d_jk}(q^{d_jk}-1);
\end{equation}
and
\begin{equation}\label{FGk*/FGk}
\frac{|(FG)^{k*}|}{|(FG)^{k}|}=\prod_{j=1}^h(1-q^{-d_jk}) ~
\mathop{\longrightarrow}_{k\to\infty}\,1
\end{equation}
with exponential convergence speed.
\end{lemma}

{\bf Proof.}~ By the assumptions, the abelian group $G$
has a subgroup $G'$ of order $m'$ and
a subgroup $G''$ of order $p^\mu$ such that $G=G''\times G'$;
hence we can assume that the group algebra $FG'$ has $h$ irreducible ideals
$E_j$ over $F$ and denote $d_j=\dim_F E_j$ for $j=1,\cdots,h$.
Then each $E_j$ is a field extension of~$F$ and
$$
 FG'=E_1\oplus E_2 \oplus\cdots\oplus E_h.
$$
Since $FG\cong FG''\otimes_F FG'$ and $FG''$ is a local ring
with head $FG''/J(FG'')\,{\cong F}$ where $J(FG'')$ denotes the Jacobson radical,
we have
\begin{equation}\label{decomp FG}
 FG\cong R_1\oplus \cdots\oplus R_h\,,
\end{equation}
where $R_j=FG''\otimes_F E_j$ for $j=1,\cdots,h$ is a local algebra with
$$
R_j/J(R_j)\cong E_j\,,\quad \dim_F R_j=p^\mu d_j\quad{\rm and}\quad
\dim_F J(R_j)=p^{\mu}d_j-d_j\,.
$$
Thus we get
\begin{equation*}
 (FG)^k\cong R_1^k\oplus \cdots\oplus R_h^k.
\end{equation*}
A vector ${\bf b}_{j}=(b_{j1},\cdots,b_{jk})$ of $R_j^k$ generates $R_j$
(i.e. $R_jb_{j1}+\cdots+R_jb_{jk}=R_j$)
if and only if the image of ${\bf b}_j$ in the
residue $R_j^k/J(R_j)^k\cong E_j^k$ is non-zero, i.e.
$R_j^{k*}=R_j^{k}\backslash J(R_j)^k$ (the difference set). So we get
$$
 |R_j^{k*}|=q^{p^{\mu}d_jk}-q^{(p^{\mu}-1)d_jk}
   =q^{(p^{\mu}-1)d_jk}(q^{d_jk}-1).
$$
It is easy to check that $(FG)^{k*}=R_1^{k*}\times\cdots\times R_h^{k*}$.
We obtain that
$$
 |(FG)^{k*}|=\prod_{j=1}^h q^{(p^{\mu}-1)d_jk}(q^{d_jk}-1);
$$
hence
$$
\frac{|(FG)^{k*}|}{|(FG)^{k}|}=\prod_{j=1}^h(1-q^{-d_jk}),
$$
which converges, as $k\to\infty$, exponentially to $1$. \qed

\section{Second moment method for the main theorem}
In this section we keep the notations in Theorem \ref{main theorem}
and Eqn (\ref{X=sum}).

In this section we always assume that $r>g_q(\delta)$ and prove that
\begin{equation}\label{aim}
\lim\limits_{n\to\infty}\Pr(X\ge 1)=1\quad
\mbox{with exponential convergence speed;}
\end{equation}
which completes the proof of Theorem 2.1, since
$\Pr\big(\Delta(C_A)\le\delta\big)=\Pr(X\ge 1)$,
see (\ref{X-Delta}), hence Eqn (\ref{aim}) implies that
$\lim\limits_{n\to\infty}\Pr\big(\Delta(C_A)>\delta\big)=0$.

By a known inequality, see \cite[Theorem 6.10]{MU}, we have that
$$
 \Pr(X\ge 1) \geq \sum_{{\bf b}\in (FG)^k}
 \frac{{\rm E}(X_{\bf b})}{{\rm E}(X|X_{\bf b}=1)},
$$
where ${\rm E}(X|X_{\bf b}=1)$ denotes the conditional expectation,
which is essentially involved in the second moment of $X$.
Such a way to investigate phase transitions (thresholds)
by means of second moments is usually named the {\em second moment method};
e.g. see \cite[Appendix]{FS}.

Since $(FG)^{k*}$ is a part of $(FG)^k$, see Eqn (\ref{I^k*}), we have
\begin{equation}\label{X>=1}
 \Pr(X\ge 1) \geq \sum_{{\bf b}\in (FG)^{k*}}
 \frac{{\rm E}(X_{\bf b})}{{\rm E}(X|X_{\bf b}=1)}\,.
\end{equation}

By the linearity of expectations, for ${\bf b_1}\in(FG)^k$ we have
$$
 {\rm E}(X|X_{\bf b_1}=1)={\rm E}\Big(\sum_{{\bf b}\in(FG)^k}
 X_{\bf b}\Big|X_{\bf b_1}=1\Big)
 =\sum_{{\bf b}\in(FG)^k}{\rm E}(X_{\bf b}|X_{\bf b_1}=1).
$$
By the conditional probability formula
(and noting that $X_{\bf b}$'s are $0$-$1$ variables), we further have
$$
{\rm E}(X_{\bf b}|X_{\bf b_1}=1)
=\frac{\Pr\big(X_{\bf b}=1\;\&\;X_{\bf b_1}=1\big)}{\Pr(X_{\bf b_1}=1)}
=\frac{{\rm E}(X_{\bf b}X_{\bf b_1})}{{\rm E}(X_{\bf b_1})}.
$$
Set
$$
{\cal A}({\bf b},{\bf b_1})=
\left\{A\in(FG)^{k\times n}\,\big|\,
  1\le {\rm w}({\bf b}A),{\rm w}({\bf b_1}A)\le mn\delta\right\},
$$
then
\begin{equation*}
{\rm E}(X_{\bf b}X_{\bf b_1})=\Pr\big(X_{\bf b}=1\;\&\; X_{\bf b_1}=1\big)
=\frac{|{\cal A}({\bf b},{\bf b_1})|}{|(FG)^{k\times n}|}
=\frac{|{\cal A}({\bf b},{\bf b_1})|}{q^{mnk}}\,.
\end{equation*}
Thus we get that
\begin{equation}\label{XXb=1}
{\rm E}(X_{\bf b}|X_{\bf b_1}=1)=
\frac{|{\cal A}({\bf b},{\bf b_1})|}{q^{mnk}{\rm E}(X_{\bf b_1})}.
\end{equation}

For any invertible $k\times k$ matrix $Q$ over $FG$,
\begin{eqnarray*}
{\cal A}({\bf b}Q,{\bf b_1}Q)
&=&\left\{A\in(FG)^{k\times n}\,\big|\,
  1\le {\rm w}({\bf b}QA),{\rm w}({\bf b_1}QA)\le mn\delta\right\}\\
&=&\left\{A\in(FG)^{k\times n}\,\big|\,
  QA\in{\cal A}({\bf b},{\bf b_1})\right\}
  =\left\{Q^{-1}A\,\big|\, A\in{\cal A}({\bf b},{\bf b_1})\right\};
\end{eqnarray*}
in particular, we have that
$|{\cal A}({\bf b}Q,{\bf b_1}Q)|=|{\cal A}({\bf b},{\bf b_1})|$.

Now we can show that
\begin{equation}\label{E|b=E|b'}
 {\rm E}(X|X_{\bf b_1}=1)={\rm E}(X|X_{\bf b_2}=1),\qquad
  \forall~~{\bf b_1},{\bf b_2}\in(FG)^{k*}.
\end{equation}
To see it, by \cite[Proposition 2.11]{FLL} we can take an invertible
$k\times k$ matrix $Q$ over $FG$ such that ${\bf b_2}={\bf b_1}Q$;
then, by Eqns (\ref{XXb=1}) and  (\ref{EXb1=EXb2}), we have
\begin{eqnarray*}
{\rm E}(X_{\bf b}|X_{\bf b_1}=1)
&=&\frac{|{\cal A}({\bf b},{\bf b_1})|}{q^{mnk}{\rm E}(X_{\bf b_1})}
=\frac{|{\cal A}({\bf b}Q,{\bf b_1}Q)|}{q^{mnk}{\rm E}(X_{\bf b_1})}\\
&=&\frac{|{\cal A}({\bf b}Q,{\bf b_2})|}{q^{mnk}{\rm E}(X_{\bf b_2})}
={\rm E}(X_{{\bf b}Q}|X_{\bf b_2}=1);
\end{eqnarray*}
hence
\begin{eqnarray*}
{\rm E}(X|X_{\bf b_1}=1)
&=&\sum_{{\bf b}\in(FG)^k}{\rm E}(X_{\bf b}|X_{\bf b_1}=1)
=\sum_{{\bf b}\in(FG)^k}{\rm E}(X_{{\bf b}Q}|X_{\bf b_2}=1);
\end{eqnarray*}
noting that ${\bf b}Q$ runs over $(FG)^k$ when ${\bf b}$ runs over $(FG)^k$,
we obtain that
$${\rm E}(X|X_{\bf b_1}=1)
=\sum_{{\bf b}\in(FG)^k}{\rm E}(X_{{\bf b}}|X_{\bf b_2}=1)
={\rm E}(X|X_{\bf b_2}=1),
$$
which is just Eqn (\ref{E|b=E|b'}).

From now on to the end of this section {\em we fix}\,
${\bf b_1}=(0,\cdots,0,1)$, which belongs obviously to $(FG)^{k*}$.
By Eqns (\ref{X>=1}), (\ref{EXb1=EXb2}) and (\ref{E|b=E|b'}), we have
$$
 \Pr(X\ge 1)\geq
 \frac{|FG^{k*}|\cdot E(X_{\bf b_1})}{E(X|X_{\bf b_1}=1)}.
$$
Thus, to prove Eqn (\ref{aim}), it is enough to prove that
\begin{equation}\label{aim'}
\lim_{n\to\infty}\frac{E(X|X_{\bf b_1}=1)}{|FG^{k*}|\cdot E(X_{\bf b_1})}
=\lim_{n\to\infty}
\frac{\sum_{{\bf b}\in(FG)^k} E(X_{\bf b}|X_{\bf b_1}=1)}
{|FG^{k*}|\cdot E(X_{\bf b_1})} = 1
\end{equation}
and it converges exponentially.

For any $A\in(FG)^{k\times n}$,
by $\underline A_i$ we denote the $i$'th row of $A$.
To compute ${\cal A}({\bf b},{\bf b_1})$, we set
$${\cal A}({\bf b_1})
=\{A\in(FG)^{k\times n}\mid 1\le{\rm w}({\bf b_1}A)\le mn\delta\};$$
since ${\bf b}=(0,\cdots,0,1)$, it is clear that
$$
{\cal A}({\bf b_1})=\left\{A\in(FG)^{k\times n}\,\big|\,
  {\bf 0}\ne\underline A_k \in\big((FG)^n\big)^{\le\delta}\right\},
$$
that is, ${\cal A}({\bf b_1})$ is the set of the
$k\times n$ matrices $A$ over $FG$ such that the $k$'th row
$\underline A_k\ne{\bf 0}$ and ${\rm w}\big(\underline A_k\big)\leq mn\delta$;
see the notation in Corollary \ref{m-weight balanced}.

Given any non-zero $(a_{k1},\cdots,a_{kn})\in\big((FG)^n\big)^{\le\delta}$,
we denote
$${\cal A}({\bf b}_1)_{(a_{k1},\cdots,a_{kn})}=\left\{
A\in(FG)^{k\times n}\,\big|\, \underline A_k=(a_{k1},\cdots,a_{kn})\right\}.$$
Then any ${\bf b}=(b_1,\cdots,b_{k-1},b_k)\in (FG)^k$ induces a map:
\begin{equation}
\begin{array}{crcl}
 \bar\beta_{{\bf b}}: & {\cal A}({\bf b_1})_{(a_{k1},\cdots.a_{kn})}
   & \longrightarrow & (FG)^n, \\
   &A & \longmapsto & {\bf b}A
   =b_1\underline A_1+\cdots+b_{k-1} \underline A_{k-1}+b_k \underline A_k;
\end{array}
\end{equation}
Set $\bar{\bf b}=(b_1,\cdots,b_{k-1})$,
$I_{\bar{\bf b}}=FGb_1+\cdots+FGb_{k-1}$ which is the
ideal of $FG$ generated by $b_1,\cdots,b_{k-1}$,
and set $d_{\bar{\bf  b}}=\dim I_{\bar{\bf b}}$.
It is easy to see that the image of the map $\bar\beta_{\bf b}$
is a coset of
$I_{\bar{\bf b}}^n\subseteq(FG)^n$ as follows
$$ I_{\bar{\bf b}}^n+b_k\underline A_k,\qquad{\rm with~ cardinality}~~
    |I_{\bar{\bf b}}^n+b_k\underline A_k|=|I_{\bar{\bf b}}^n|
     =q^{d_{\bar{\bf b}}n};$$
and the number of the pre-images in
${\cal A}({\bf b_1})_{(a_{k1},\cdots.a_{kn})}$ of
any ${\bf a}\in I_{\bar{\bf b}}^n+b_k\underline A_k$ is equal to
$\frac{q^{mn(k-1)}}{q^{d_{\bar{\bf b}}n}}=q^{mn(k-1)-d_{\bar{\bf b}}n}$,
 which is independent of the choices of $\bf a$ and $(a_{k1},\cdots,a_{kn})$.
Thus the cardinality of the pre-image in
${\cal A}({\bf b_1})_{(a_{k1},\cdots,a_{kn})}$ of the set
$(I_{\bar{\bf b}}+b_k \underline A_k)^{\le\delta}\backslash\{{\bf 0}\}$ is
$$
\big(|(I_{\bar{\bf b}}^n+b_k \underline A_k)^{\le\delta}|-\lambda\big)
q^{mn(k-1)-d_{\bar{\bf b}}n}
$$
with
\begin{equation}\label{lambda}
\lambda=\begin{cases}1, & {\rm if}~{\bf 0}\in I_{\bar{\bf b}}+b_k \underline A_k;\\
 0, &\mbox{otherwise;}\end{cases}
\end{equation}
hence
$$
 |{\cal A}({\bf b},{\bf b_1})| =
 \sum_{{\bf 0}\ne(a_{k1},\cdots,a_{kn})\in((FG)^n)^{\le\delta}}
\big(|(I_{\bar{\bf b}}^n+b_k \underline A_k)^{\le\delta}|-\lambda\big)
q^{mn(k-1)-d_{\bar{\bf b}}n};
$$
that is
$$
 |{\cal A}({\bf b},{\bf b_1})| = \left(|((FG)^n)^{\le\delta}|-1\right)
 \left( |(I_{\bar{\bf b}}^n+b_k \underline A_k)^{\le\delta}|-\lambda\right)\cdot
 q^{mn(k-1)-d_{\bar{\bf b}}n}.
$$
But $|((FG)^n)^{\le\delta}|-1=q^{mn}{\rm E}(X_{\bf b_1})$,
see Eqn (\ref{EXb*}). By Eqn (\ref{XXb=1}), we get that
\begin{equation}\label{EXbb1}
 {\rm E}(X_{\bf b}|X_{\bf b_1}=1) =\left(
 |(I_{\bar{\bf b}}^n+b_k \underline A_k)^{\le\delta}|-\lambda\right)
 \cdot q^{-d_{\bar{\bf b}}n}.
\end{equation}

By the disjoint union $(FG)^{k-1}=\bigcup_{I\le FG}I^{(k-1)*}$ again,
cf. Eqn (\ref{decomp EXI})
(but this time we consider $\bar{\bf b}$ which has length $k-1$), we have
\begin{eqnarray*}
\sum_{{\bf b}\in(FG)^k}{\rm E}(X_{\bf b}|X_{\bf b_1}=1)
=\sum_{I\le FG}\,\sum_{\bar{\bf b}\in I^{(k-1)*}}\,\sum_{b_k\in FG}
{\rm E}(X_{\bf b}|X_{\bf b_1}=1),
\end{eqnarray*}
where $\bar{\bf b}=(b_1,\cdots,b_{k-1})$
and ${\bf b}=(b_1,\cdots,b_{k-1},b_k)$.
Thus
\begin{equation}\label{S_I}
\frac{\sum_{{\bf b}\in(FG)^k} {\rm E}(X_{\bf b}|X_{\bf b_1}=1)}
{|(FG)^{k*}|\cdot {\rm E}(X_{\bf b_1})}
=\sum_{0\ne I\le FG} S_I
\end{equation}
with
$$
S_I=\frac{\sum_{\bar{\bf b}\in I^{(k-1)*}}\,\sum_{b_k\in FG}
{\rm E}(X_{\bf b}|X_{\bf b_1}=1)}
{|(FG)^{k*}|\cdot{\rm E}(X_{\bf b_1})}.
$$
We compute the $S_I$'s for ideals $I$ of $FG$ into two cases.

{\bf Case 1.}~ $0\ne I\ne FG$; note that
there are only finitely many such ideals of $FG$.
Let $d_I=\dim I$; then $d_I<m$.
By Remark \ref{g-code balanced} and Corollary \ref{product balanced},
we see that $I_{\bar{\bf b}}^n+b_k\underline A_k$ is a balanced code and
$$|(I_{\bar{\bf b}}^n+b_k \underline A_k)^{\le\delta}|-\lambda\le
 |(I_{\bar{\bf b}}^n+b_k \underline A_k)^{\le\delta}|
  \leq q^{d_{\bar{\bf b}}nh_q(\delta)}.$$
By Eqn (\ref{EXbb1}) and the above inequality, we obtain that
\begin{equation*}
E(X_{\bf b}|X_{\bf b_1}=1)\le q^{-d_{\bar{\bf b}}n g_q(\delta)}.
\end{equation*}
Note that $d_{\bar{\bf b}}=d_I$ for any ${\bar{\bf b}}\in I^{(k-1)*}$,
$|I^{(k-1)*}|\le |I^{(k-1)}|=q^{d_I(k-1)}$ and $|FG|=q^m$.
By the inequality (\ref{k*E}) and the above inequality, we have
\begin{eqnarray*}
S_I&\le&
\frac{q^{d_I(k-1)}q^m\cdot q^{-d_Ing_q(\delta)}}
 {\frac{|(FG)^{k*}|}{|(FG)^{k}|}
  \big(q^{mn(\frac{k}{n}-g_q(\delta))-\frac{1}{2}\log_q(mn)}
     -q^{-mn(1-\frac{k}{n})}\big)}\\
&=&\frac{|(FG)^{k}|}{|(FG)^{k*}|}\cdot
\frac{q^{d_In(\frac{k}{n}-g_q(\delta))-d_I+m}}
 {q^{mn(\frac{k}{n}-g_q(\delta))-\frac{1}{2}\log_q(mn)}
   -q^{-mn(1-\frac{k}{n})}}\,.
\end{eqnarray*}
Since $k=[rn]$ and $1>r>g_q(\delta)$, there is a real number
$\gamma>0$ such that for large enough $n$ we have
$\frac{k}{n}-g_q(\delta)>\gamma$ and $1-\frac{k}{n}>\gamma$; hence
\begin{equation*}
\frac{1}{S_I}\ge\frac{|(FG)^{k*}|}{|(FG)^{k}|}\big(
q^{\gamma(m-d_I)n+\frac{1}{2}\log_q(mn)+d_I-m}-O(q^{-\gamma n})\big),
\end{equation*}
where $O(q^{-\gamma n})$ stands for a quantity bounded from above by a
multiple of $q^{-\gamma n}$. Recalling that $d_I<m$ and
$\lim\limits_{n\to\infty}\frac{|(FG)^{k}|}{|(FG)^{k*}|}=1$
(see Eqn (\ref{FGk*/FGk})), we get that
\begin{equation}\label{S=0}
\lim_{n\to\infty}S_I =0,\qquad {\rm if}~~ I\ne FG,
\end{equation}
and the limit converges exponentially.

{\bf Case 2.}~ $I=FG$, and $\bar{\bf b}\in I^{(k-1)*}$.
Then $d_{\bar{\bf b}}=m$, and $I_{\bar{\bf b}}^n=(FG)^n$,
hence $I_{\bar{\bf b}}^n+b_k\underline A_k=(FG)^n$;
in particular, $\lambda=1$ in Eqn (\ref{lambda}).
By Eqn (\ref{EXb*}), we get
$$|(I_{\bar{\bf b}}^n+b_k\underline A_k)^{\le\delta}|-\lambda
=|((FG)^n)^{\le\delta}|-1=q^{mn}{\rm E}(X_{\bf b_1}).$$
By Eqn (\ref{EXbb1}) we can compute
\begin{eqnarray*}
S_I&=&\frac{|(FG)^{(k-1)*}|\cdot q^m\cdot q^{mn}{\rm E}(X_{\bf b_1})
\cdot q^{-mn}}  {|(FG)^{k*}|\cdot{\rm E}(X_{\bf b_1})}\\
&=&\frac{|(FG)^{(k-1)*}|\cdot q^m}{|(FG)^{k*}|}
=\frac{|(FG)^{(k-1)*}|}{|(FG)^{k-1}|}\cdot\frac{|(FG)^{k}|}{|(FG)^{k*}|}\,.
\end{eqnarray*}
By the exponential convergence
$\lim\limits_{n\to\infty}\frac{|(FG)^{k}|}{|(FG)^{k*}|}=1$
(see Eqn (\ref{FGk*/FGk})) again, we get the following exponential convergent limit:
\begin{equation}\label{S=1}
\lim_{n\to\infty}S_I =1,\qquad {\rm if}~~ I=FG.
\end{equation}

Finally, by Eqn (\ref{S_I}), Eqn (\ref{S=0}) and Eqn (\ref{S=1}), we obtain that
$$
\lim_{n\to\infty}\frac{\sum_{{\bf b}\in(FG)^k} E(X_{\bf b}|X_{\bf b_1}=1)}
{|(FG)^{k*}|\cdot E(X_{\bf b_1})} =1
$$
and it converges exponentially; this is just what Eqn (\ref{aim'}) requires.

\section{Random quasi-abelian codes}

Keep notations in (\ref{parameters}), (\ref{ensemble}) and (\ref{C_A}).

Recall that for $A\in(FG)^{k\times n}$
the rate $R(C_A)=\frac{k}{n}$ if and only if
the $FG$-rank of $A$ is equal to $k$;
at that case we say that {\em $A$ is full-rank}.

In order to get random quasi-abelian codes of rate $\frac{k}{n}\approx r$,
we consider the probability space ${\cal F}$,
which sample space is
$\{A\in(FG)^{k\times n}\mid \mbox{$A$ is full-rank}\}$
and probability function is equiprobability.
Take ${\cal A}\in{\cal F}$, construct
$C_{\cal A}$ the same as in (\ref{C_A}).
By $\Pr_{\cal F}\big(\Delta(C_{\cal A})>\delta\big)$
we emphasize that the probability is computed over the
probability space ${\cal F}$.

\begin{corollary} Let notation be as above. Then
$$
\lim\limits_{n\to\infty}{\Pr}_{\cal F}\big(\Delta(C_{\cal A})>\delta\big)
  =\begin{cases}1, & r<g_q(\delta);\\ 0, & r>g_q(\delta); \end{cases}
$$
and both the limits converge exponentially.
\end{corollary}

{\bf Proof.}~ Let $A\in(FG)^{k\times n}$.
By the total probability formula we have
$$\begin{array}{r}
\Pr(\Delta(C_A)>\delta)=
\Pr\big(\Delta(C_A)>\delta\,|\,\mbox{$A$ is full-rank})
  \cdot\Pr(\mbox{$A$ is full-rank})~+~\qquad\\
\Pr\big(\Delta(C_A)>\delta\,|\,\mbox{$A$ is not full-rank}\big)
  \cdot\Pr(\mbox{$A$ is not full-rank}).
\end{array}$$
Noting that
$$\textstyle
\Pr\big(\Delta(C_A)>\delta\,|\,\mbox{$A$ is full-rank}\big)
  =\Pr_{\cal F}\big(\Delta(C_{\cal A})>\delta\big);
$$
and by Lemma \ref{full-rank} below,
$$
 \lim_{n\to\infty}\Pr(\mbox{$A$ is not full-rank})=0,\quad
 \lim_{n\to\infty}\Pr(\mbox{$A$ is full-rank})=1\,;
$$
so we get
$$\textstyle
\lim\limits_{n\to\infty}\Pr\big(\Delta(C_A)>\delta\big)=
 \lim\limits_{n\to\infty}\Pr_{\cal F}\big(\Delta(C_{\cal A})>\delta\big).
$$
Then the corollary follows from Theorem \ref{main theorem} at once.
\qed

\medskip
From the first part (the case $r<g_q(\delta)$) of the above corollary we obtain the following result immediately.

\begin{corollary}
For any $(r,\delta)\in(0,1)\times(0,1-q^{-1})$ satisfying
that $r<g_q(\delta)$, there exists a series of quasi-$FG$ codes
$C_1,C_2,\cdots$ such that:

(i) ~ the length of $C_i$ goes to infinity;

(ii)~ $\lim\limits_{i\to\infty} R(C_i) = r$;

(iii)~ $\lim\limits_{i\to\infty} \Delta(C_i)\ge\delta$.
\end{corollary}

\begin{lemma}\label{full-rank} Let $A\in(FG)^{k\times n}$ where
$k=[rn]$ and $0<r<1$. Then
$\lim\limits_{n\to\infty}
  \Pr\big(\mbox{\rm $A$ is full-rank}\big)=1$, or equivalently,
$\lim\limits_{n\to\infty}
  \Pr\big(\mbox{\rm $A$ is not full-rank}\big)=0$;
and both the limits converge exponentially.
\end{lemma}

{\bf Proof.}~ Let $A=\big(a_{\alpha\beta}\big)_{k\times n}$
with $a_{\alpha\beta}\in FG$.
We quote the decomposition of $FG$ in (\ref{decomp FG}) and adopt their notations.
Each $a_{\alpha\beta}$ can be written as
$$
 a_{\alpha\beta}=\big(a_{\alpha\beta}^{(1)},\cdots,a_{\alpha\beta}^{(h)}\big),
 \qquad a_{\alpha\beta}^{(j)}\in R_j;
$$
hence the matrix $A$ can be rewritten as
$A=\big(A^{(1)},\cdots,A^{(h)}\big)$ with
$A^{(j)}=\big(a_{\alpha\beta}^{(j)}\big)_{k\times n}$ being
$k\times n$ matrix over the local algebra $R_j$
(cf. \cite[Eqn (2.2)]{FLL}), and $A$ is full-rank
if and only if every $A^{(j)}$ is full-rank over
the field $E_j$ for $j=1,\cdots,h$, cf. \cite[Lemma 2.2]{FLL}.
For $1\le i\ne j\le h$, it is clear that
$A^{(i)}\in R_i^{k\times n}$ and $A^{(j)}\in R_j^{k\times n}$
are randomly independent of each other. So we have
\begin{equation}\label{independent-Pr}
\Pr\big(\mbox{\rm $A$ is full-rank}\big)=
\prod_{j=1}^h \Pr\big(\mbox{\rm $A^{(j)}$ is full-rank}\big).
\end{equation}

Let $j$ with $1\le j\le h$ be given. First we claim that
\begin{equation}\label{A barA}
 \mbox{$A^{(j)}$ is full-rank}~~ \iff~~ \mbox{$\bar A^{(j)}$ is full-rank}\,,
\end{equation}
where $\bar A^{(j)}=\big(\bar a_{\alpha\beta}^{(j)}\big)_{k\times n}$
is the image of $A^{(j)}$ in $E_j^{k\times n}$, i.e.
each $\bar a_{\alpha\beta}^{(j)}$ is the image of the element
$a_{\alpha\beta}^{(j)}$ in the residue filed $E_j=R_j/J(R_j)$.
To see it, we remark that  $A^{(j)}$ is full-rank if and only if it is right invertible,
cf. \cite[Lemma 2.6]{FLL}. Suppose that $A^{(j)}$ is full-rank,
then $A^{(j)}B=I_{k\times k}$ for a $B\in R_j^{n\times k}$,
where $I_{k\times k}$ stands for the identity $k\times k$ matrix;
mapping them to matrices over $E_j$,
we get that $\bar A^{(j)}\bar B=\bar I_{k\times k}$, which implies that
$\bar A^{(j)}$ is full-rank over $E_j$.
Conversely, if $\bar A^{(j)}$ is full-rank over $E_j$, then
$\bar A^{(j)}\bar B=\bar I_{k\times k}$ for a $B\in R_j^{n\times k}$, hence
$$
 A^{(j)}B=I_{k\times k}+C, \qquad
  {\rm with}\quad C\in J(R_j)^{k\times k};
$$
since $C$ is a nilpotent matrix, $I_{k\times k}+C$ is an invertible matrix;
hence $A^{(j)}$ is full-rank over $R_j$.

Next we claim that
\begin{equation}\label{Pr(A)}
\Pr\big(\mbox{\rm $A^{(j)}$ is not full-rank}\big)
  \le q^{d_j(k-n)}\approx q^{d_j(r-1)n}.
\end{equation}
To see it, we note three points:  $\bar A^{(j)}$ is not full-rank if and only if
there a $(k-1)$-dimensional subspace of $E_j^k$ which contains all
the columns of $\bar A^{(j)}$;
the probability that a $(k-1)$-dimensional subspace of $E_j^k$
contains all the columns of $\bar A^{(j)}$ is $1/q^{d_jn}$
(recall that the cardinality $|E_j|=q^{d_j}$);
the number of the $(k-1)$-dimensional subspaces of $E_j^k$ is
$\frac{q^{d_jk}-1}{q^{d_j}-1}\le q^{d_jk}$;
thus
$$
 \Pr\big(\mbox{\rm $\bar A^{(j)}$ is not full-rank}\big)\le
 q^{d_jk}\cdot\frac{1}{q^{d_jn}}=q^{d_j(k-n)}.
$$
Each matrix in $E_j^{k\times n}$ has exactly $|J(R_j)|^{kn}$ inverse images in
$R_j^{k\times n}$. So the claim~(\ref{Pr(A)}) follows from
the above inequality and the conclusion (\ref{A barA}).

Finally, since $r-1<0$, from the inequality (\ref{Pr(A)}) we obtain
$$
\lim\limits_{n\to\infty}\Pr\big(\mbox{\rm $A^{(j)}$ is not full-rank}\big)
\le\lim\limits_{n\to\infty}q^{d_j(r-1)n}=0.
$$
By Eqn (\ref{independent-Pr}), we are done for the lemma.
\qed

\section*{Acknowledgements}
The research of the authors is supported by NSFC
with grant numbers 11171370 and 11271005.

\end{document}